\documentclass[preprint,showpacs,preprintnumbers,amsmath,amssymb]{revtex4}

\usepackage{graphicx}
\usepackage{dcolumn}
\usepackage{bm}

\begin{document}
\title{Quantum phase transitions in coupled two-level atoms in a single-mode cavity}
\author{Qing-Hu Chen$^{1,2,*}$, Tao Liu$^{3}$, Yu-Yu Zhang$^{2}$, and Ke-Lin Wang$%
^{4}$}
\date{\today}

\address{
$^{1}$ Center for Statistical and Theoretical Condensed Matter
Physics, Zhejiang Normal University, Jinhua 321004, P. R. China  \\
$^{2}$ Department of Physics, Zhejiang University, Hangzhou 310027,
P. R. China \\
$^{3}$ Department of Physics, Southwest University of  Science and Technology, Mianyang 621010, P.  R.  China\\
$^{4}$ Department of Modern Physics, University of  Science and
Technology of China,  Hefei 230026, P.  R.  China
 }
\begin{abstract}
The dipole-coupled two-level atoms(qubits) in   a single-mode
resonant cavity is studied by extended bosonic coherent states. The
numerically exact solution is presented. For finite systems, the
first-order quantum phase transitions  occur at the strong
interatomic interaction. Similar to the original Dicke model, this
system exhibits  a second-order quantum phase transition from the
normal to the superradiant phases. Finite-size scaling for several
observables, such as  the average fidelity susceptibility, the order
parameter, and concurrence are performed for different interatomic
interactions. The obtained scaling exponents suggest that
interatomic interactions do not change the universality class.
\end{abstract}

 \maketitle

\draft

\section{introduction}

The coherent emission from underdamped one-dimensional Josephson
arrays coupled to a single-mode electromagnetic cavity has been
experimentally studied\cite{jjarys}.  This system resembles that of
the Dicke model\cite{dicke}, which describes N two-level atoms
(qubits) coupled to a cavity field.  It has been shown\cite{dickexy}
that a modified Dicke Hamiltonian including a dipole-dipole
interaction between the junctions can describe the cavity-junction
system better  than  the original Dicke model.

The modified Dicke Hamiltonian  is also an extension of effective
two-qubit model\cite{atoms}. A small number of coupled qubits  in a
single-mode cavity is also very interesting, because it may be
realized in several solid-state systems, such as an ensemble of
quantum dots \cite{Scheibner}, Bose-Einstein condensates
\cite{Schneble}, coupled arrays of optical cavities  used to
simulate and study the behavior of strongly correlated
systems\cite{Hartmann}, and the superconducting quantum interference
device coupled with a nanomechanical resonator\cite{Wallraff,squid}.

The original Dicke model without a rotating-wave approximation (RWA)
on the large scale was  exactly solved by the present authors in the
numerically sense\cite{chenqh}. Most works on the modified Dicke
model are limited to the  rotating-wave approximation. With the
progress of the fabrication, the artificial atoms may interact very
strongly with on-chip resonant
circuits\cite{Wallraff,squid,Peropadre,exp}, the rotating-wave
approximation can not describe well the strong coupling
regime\cite{theory}, so the numerically exact solution to the
modified Dicke model without rotating-wave approximation is also of
considerable significance and highly called for.

Quantum phase transitions (QPTs) in the original Dicke model has
attracted considerable attentions
recently\cite{Emary,Lambert1,Lambert2,liberti,vidal,reslen,plastina}.
With the consideration of additional interatomic interactions, one
nature question is its effect on the modified Dicke model.

In this paper, we extend our  previous  exact technique to solve the
finite size modified Dicke model where dipole-dipole interaction
between the qubits are taken into account. The QPTs are then studied
systematically. The paper is organized as follows. In Sec.II, the
numerically exact solution to the finite-size modified Dicke model
is proposed in detail, and the analytical solution in the
thermodynamic limit is also presented. The numerical results for
both small and large system size are given in Sec.III, where the
characterization of the QPTs  also also performed. The brief summary
is presented finally in the last section.

\section{Model Hamiltonian}

dipole-coupled $N$ two-level atoms interacting with a single-mode
cavity\cite{jjarys,atoms} can be described by the following modified
Dicke model\cite{dickexy}
\begin{equation}
H=\omega a^{+}a+\frac \Delta 2\sum_{i=1}^N\sigma _z^i+\frac \lambda {\sqrt{N}%
}\left( a^{+}+a\right) \sum_{i=1}^N\sigma _x^i+\frac \Omega {2N}\sum_{i\neq
j}^N\left( \sigma _x^i\sigma _x^j+\sigma _y^i\sigma _y^j\right) ,
\end{equation}
where $a^{+}$ and $a$ are the field annihilation and creation operators, $%
\Delta $ and $\omega $ are the transition frequency of the qubit and
the frequency of the single bosonic mode, $\lambda $ is the coupling
constant of the atom and cavity, $\Omega$ is the interacting
strength of two two-level atoms, and $\sigma _k^i(k=x,y,z)$ is the
Pauli matrix of the $i$ junction. For convenience, the Hamiltonian
can be rewritten in terms of the collective spin operators:
$S_k=\sum_{i=1}^N\sigma _k^i/2,\;\;(k=x,y,z)$ and $S_{\pm }=S_x\pm
iS_y$
\begin{equation}
H=\omega a^{+}a+\Delta S_z+\frac{2\lambda }{\sqrt{N}}\left( a^{+}+a\right)
S_x+\frac{2\Omega }N(\mathbf{S}^2-S_z^2),
\end{equation}
Where a constant is neglected. Without atom-cavity coupling, i.e.$\lambda =0,
$ we have
\begin{equation}
H_{atoms}=\Delta S_z+\frac{2\Omega }N(\mathbf{S}^2-S_z^2),
\end{equation}
Note that the Dicke state $\{\left| j,m\right\rangle ,m=-j,-j+1,...j-1,j\}$
is the eigenstate of $S^2$ and $S_z$ with the eigenvalues $j(j+1)$ and $m.$
The eigen energy of $H_{atoms}$ is given by
\begin{equation}
E=\Delta m+\frac{2\Omega }N(j(j+1)-m^2),
\end{equation}
The ground-state energy is easily obtained as
\[
E_0=\left( -\Delta +\frac{2\Omega }N\right) j
\]
So in the ground-state, we have $j=0$ for $\frac{2\Omega }N>\Delta $ and $%
j=N/2$ otherwise. Actually, Eq. (3) is just the Hamiltonian of the
isotropic Lipkin-Meshkov-Glick model\cite{LMG}, where QPT is of the
first-order, as is clearly shown above.

Next, we use a transformed Hamiltonian with a rotation around an $y$
axis by an angle $\frac \pi 2$ , so $S_x\rightarrow
S_z,S_z\rightarrow -S_x$ the Hamiltonian now reads
\begin{equation}
H^{\prime }=\omega a^{+}a-\frac \Delta 2(S_{+}+S_{-})+\frac{2\lambda }{\sqrt{%
N}}\left( a^{+}+a\right) S_z+\frac{2\Omega }N(\mathbf{S}^2-S_x^2),
\end{equation}
where $S_{+}$ and $S_{-}$ are the collective spin raising and lowing
operators and obey the SU(2) Lie algebra
$[S_{+},S_{-}]=2S_z,[S_z,S_{\pm }]=\pm S_{\pm }$. Since $[H^{\prime
},S^2]=0,$ we suppose $S^2=j(j+1)$, where the total angular momentum
$j=N/2-r$.  Due to the interaction between any two-level atoms,
similar to the above discussion for $\lambda =0$, we have no reason to set $%
r=0$, unlike in the Dicke model\cite{chenqh}. In the practical
calculations for the ground-state, $r$ is regarded as a variational
integer number and will be determined by the minimization of the
ground-state energy. So the Hilbert space of this algebra is spanned
by the Dicke state $\{\left| j,m\right\rangle
,m=-j,-j+1,...j-1,j\}$, which is the eigenstate of $S^2$ and $S_z$
with the eigenvalues $j(j+1)$ and $m$.

The Hilbert space of the total system can be expressed in terms of the basis
$\{\left| \varphi _m\right\rangle _b\bigotimes \left| j,m\right\rangle \}$,
where the state of bosons$\left| \varphi _m\right\rangle _b$ and the integer
$j$ are to be determined. A ''natural'' basis for bosons is Fock state $%
\left| l\right\rangle _b=[(a^{+})^l/\sqrt{l!}]\left| 0\right\rangle
_b $. As in the Dicke model, the bosonic number here is not
conserved either, so the bosonic Fock space has infinite dimensions,
the standard diagonalization procedure (see, for example, Ref.
\cite{Emary} ) is to apply a truncation procedure considering only a
truncated number of bosons. Typically, the convergence is assumed to
be achieved if the ground-state energy is determined within a very
small relative errors. Within this method, one has to diagonalize
very large, sparse Hamiltonian in strong coupling regime and/or in
adiabatic regime. Furthermore, the calculation becomes prohibitive
for larger system size since the convergence of the ground-state
energy is very slow. Interestingly, this problem can be circumvented
in the following procedure.

By the displacement transformation $A_m=a+g_m$ with $g_m=2\lambda m/\omega
\sqrt{N}$, the Schr$\stackrel{..}{o}$ dinger equation can be described in
columnar matrix, and its $m$ row reads
\begin{eqnarray}
&&-\Delta j_m^{-}\left| \varphi _m\right\rangle _b\left| j,m-1\right\rangle
-\Delta j_m^{+}\left| \varphi _m\right\rangle _b\left| j,m+1\right\rangle
\nonumber \\
&&-\frac{2\Omega }Nj_m^{-}j_{m-1}^{-}\left| \varphi _m\right\rangle _b\left|
j,m-2\right\rangle -\frac{2\Omega }Nj_m^{+}j_{m+1}^{+}\left| \varphi
_m\right\rangle _b\left| j,m+2\right\rangle  \nonumber \\
&&+\omega \left( A_m^{+}A_m-g_m^2\right) \left| \varphi
_m\right\rangle _b\left| j,m\right\rangle +\frac 2\Omega N\left[
j(j+1)-(j_{m-1}^{+}j_{m}^{-}+j_{m+1}^{-}j_{m}^{+})\right] \left|
\varphi _m\right\rangle _b\left| j,m\right\rangle  \nonumber \\
&=&E\left| \varphi _m\right\rangle _b\left| j,m\right\rangle ,
\end{eqnarray}
where $j_m^{\pm }=\frac 12\sqrt{(j(j+1)-m(m\pm 1)}$. Left multiplying $%
\{\left\langle n,j\right| \}$ gives a set of equations
\begin{eqnarray}
&&-\Delta j_n^{-}\left| \varphi _{n-1}\right\rangle _b-\Delta j_n^{+}\left|
\varphi _{n+1}\right\rangle _b-\frac{2\Omega }Nj_n^{-}j_{n-1}^{-}\left|
\varphi _{n-2}\right\rangle _b-\frac{2\Omega }Nj_n^{+}j_{n+1}^{+}\left|
\varphi _{n+2}\right\rangle _b  \nonumber \\
&&+\omega \left( A_n^{+}A_n-g_n^2\right) \left| \varphi
_n\right\rangle _b+\frac 2\Omega N\left[
j(j+1)-(j_{n-1}^{+}j_{n}^{-}+j_{n+1}^{-}j_{n}^{+})\right] \left|
\varphi _n\right\rangle _b=E\left| \varphi _n\right\rangle _b,
\end{eqnarray}
where $n=-j,-j+1,...j$.

Note that the linear term for the bosonic operator $a(a^{+})$ is removed,
and a new free bosonic field with operator $A(A^{+})$ appears. In the next
step, we naturally choose the basis in terms of this new operator, instead
of $a(a^{+})$, by which the bosonic state can be expanded as
\begin{eqnarray}
\left| \varphi _n\right\rangle _b
&=&\sum_{k=0}^{N_{tr}}c_{n,k}(A_n^{+})^k\left| 0\right\rangle _{A_n}
\nonumber \\
&=&\sum_{k=0}^{N_{tr}}c_{n,k}\frac 1{\sqrt{k!}%
}(a^{+}+g_n)^ke^{-g_na^{+}-g_n^2/2}\left| 0\right\rangle _a,  \label{ }
\end{eqnarray}
where $N_{tr}$ is the truncated bosonic number in the Fock space of $A(A^{+})
$. As we know that the vacuum state $\left| 0\right\rangle _{A_n}$ is just a
bosonic coherent-state in $a(a^{+})$ with an eigenvalue $g_n$\cite{chenqh}.
So this new basis is overcomplete, and actually does not involve any
truncation in the Fock space of $a(a^{+})$, which highlights the present
approach. It is also clear that many-body correlations for bosons are
essentially included in extended coherent states (5). Left multiplying state
$_{A_n}\left\langle l\right| $ yields
\begin{eqnarray}
&&\left\{ \omega (l-g_n^2)+\frac 2\Omega N\left[
j(j+1)-(j_{n-1}^{+}j_{n}^{-}+j_{n+1}^{-}j_{n}^{+})\right] \right\}
c_{n,l}\nonumber \\
&&-\Delta \sum_{k=0}^{N_{tr}}\left( j_n^{-}\text{ }_{A_n}\langle
l\left| k\right\rangle _{A_{n-1}}c_{n-1,k}+j_n^{+}\text{
}_{A_n}\langle
l\left| k\right\rangle _{A_{n+1}}c_{n+1,k}\right)   \nonumber \\
&&-\frac{2\Omega }N\sum_{k=0}^{N_{tr}}\left(
j_n^{-}j_{n-1}^{-}\;_{A_n}\langle l\left| k\right\rangle
_{A_{n-2}}c_{n-2,k}+j_n^{+}j_{n+1}^{+}\;_{A_n}\langle l\left|
k\right\rangle _{A_{n+2}}c_{n+2,k}\right) =Ec_{n,l},
\end{eqnarray}
where
\begin{eqnarray}
_{A_n}\langle l\left| k\right\rangle _{A_{n-1}}
&=&(-1)^lD_{l.k}(G),\;\;\;_{A_n}\langle l\left| k\right\rangle
_{A_{n+1}}=(-1)^kD_{l.k}(G),  \nonumber \\
_{A_n}\langle l\left| k\right\rangle _{A_{n-2}}
&=&(-1)^lD_{l.k}(2G),\;\;\;_{A_n}\langle l\left| k\right\rangle
_{A_{n+2}}=(-1)^kD_{l.k}(2G),
\end{eqnarray}
with
\[
D_{l,k}=e^{-G^2/2}\sum_{r=0}^{min[l,k]}\frac{(-1)^{-r}\sqrt{l!k!}G^{l+k-2r}}{%
(l-r)!(k-r)!r!},\;G=\frac{2\lambda }{\omega \sqrt{N}}
\]
Eq. (6) is just a eigenvalue problem, which can be solved by the
exact Lanczos diagonalization approach in dimensions
$(2j+1)(N_{tr}+1)$\cite {chenqh} . Note that the eigenvalue problem
in the pure Dicke model can be reproduced if set $\Omega =0$. As
before, to obtain the true exact results, in principle, the
truncated number $N_{tr}$ should be taken to infinity. Fortunately,
it is not necessary. It is found that finite terms in state (5) are
sufficient to give very accurate results with a relative errors less
than $10^{-6}$ in the whole parameter space. We believe that we have
exactly solved this model numerically.

\textsl{Thermodynamic limit}.-- In order to study the quantum phase
transition explicitly of this model, we should evaluate the
transition point in the thermodynamical limit $N\rightarrow \infty
$. First, we use the Holstein-Primakoff transition~ of the
collective angular momentum operators
defined as $S_{+}=a^{\dagger }\sqrt{N-a^{\dagger }a}$, $S_{-}=\sqrt{%
N-a^{\dagger }a}a$, and $S_z=a^{\dagger }a-N/2$, where $[a,a^{\dagger }]=1$.
Second, we introduce shifting boson operators $c^{\dagger }$ and $c^{\dagger
}$ with properly scaled auxiliary parameters $\alpha $ and $\beta $ such
that $c^{\dagger }=b^{\dagger }+\sqrt{N}\alpha $ and $d^{\dagger
}=a^{\dagger }-\sqrt{N}\beta $ to describe the collective behavior of the
Hamiltonian in Eq.(~1). Finally, by means of the boson expansion approach,
we expand the $H$ with respect to the new operators $c^{\dagger }$ and $%
d^{\dagger }$ as power series in $1/N$. According to Hamiltonian (2), we
have the scaled ground state energy
\[
\frac{E_0(\alpha ,\beta )}N=\omega \alpha ^2-4\lambda \alpha \beta \sqrt{%
1-\beta ^2}+\Delta (\beta ^2-\frac 12)-2\Omega (\beta ^2-\frac 12)^2+\frac
\Omega 2
\]
The critical points can be determined from the equilibrium condition $%
\partial [E_0(\alpha ,\beta )/N]/\partial \alpha =0$ and $\partial
[E_0(\alpha ,\beta )/N]/\partial \beta =0$, which leads to two equations
\begin{eqnarray}
\omega \alpha -2\lambda \beta \sqrt{1-\beta ^2} &=&0,  \nonumber \\
2\alpha \lambda \sqrt{1-\beta ^2}-2\alpha \lambda \beta ^2\frac 1{\sqrt{%
1-\beta ^2}}-\beta \Delta +4\Omega \beta (\beta ^2-\frac 12) &=&0.
\end{eqnarray}
Then we can obtain the critical atom-cavity coupling constant in the
second-order QPT $\lambda _c=\sqrt{\omega (\Delta +2\Omega )}/2$

\section{Results and discussions}

The exactly numerical results are presented in this section to study
the properties of QPT in this model. Without loss of generality, we
set $\omega=\Delta$ in the whole calculation.  The units are taken
of $\omega=1$ for convenience.

\begin{figure}[tbp]
\centering
\includegraphics[width=10cm]{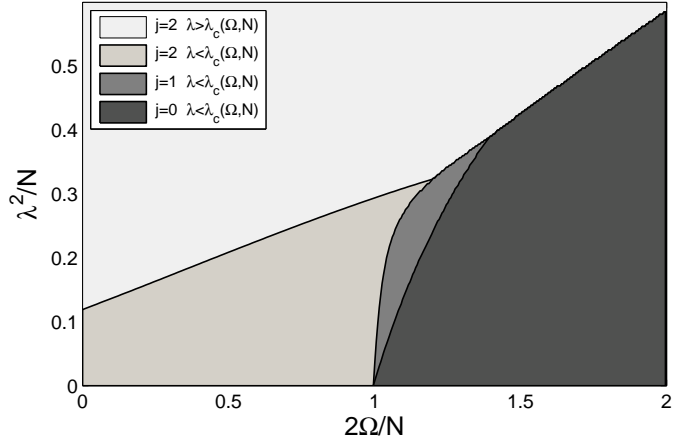}
\caption{The total momentum angular in the ground-state $j$ as a
function of the coupling constant $\lambda$ and the interaction
strength $\Omega$ between atoms for $N=4$. } \label{firstorder}
\end{figure}

First, we study a finite number of coupled two-level atoms in a
cavity. The maximum value of $j$ is $j_{\max }=N/2$. It is very
surprising that $j\neq j_{\max }$ in the ground-state for finite
systems, unlike in the pure Dicke model. In Fig. \ref{firstorder}(a)
and (b), we plot the value of j as a function of $\lambda $ and
$\Omega ^{\prime }=2\Omega /N$ in the ground-state for $N=4$. In the
weak interaction of the two-level atoms with small $\Omega ^{\prime
}$, $j=j_{max }$,  similar to the pure dicke model where the
interaction of the two- level atoms is neglected. As $\Omega
^{\prime }$ increases, the value of $j$ is reduced. In the
intermediate coupling range of atom-cavity $\lambda ,$ the value of
$j$ is reduced gradually by step $1$, until to zero in the strong
atomic interaction. This phenomena disappears in the strong coupling
range of atom-cavity $\lambda$,  where $j$ is always equal to
$j_{max }$. For $\lambda=0$, according to Eq. (4), the value of $j$
jumps to zero when the $2\Omega /N = 1$, consistent with the
observation in Fig. \ref{firstorder}.  The above observations hold
true for any finite number of  atoms.

\begin{figure}[tbp]
\begin{center}
\includegraphics[width=12cm]{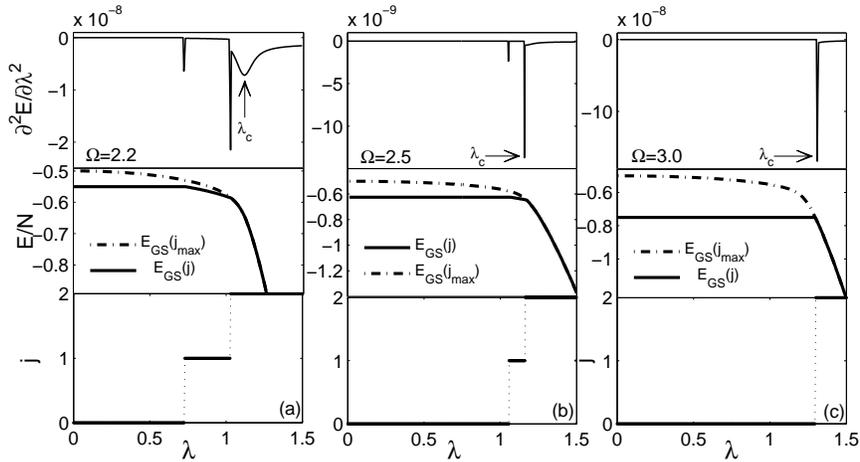}
\vspace{-0.8cm} \caption{The average ground-state energy $E/N$,
its second derivatives $\partial^2 E/\partial \lambda^2$, and the
total angular momentum $j$  as a function of $\lambda$ for
different atomic interactions $\Omega=2.2$ (a), $\Omega=2.5$ (b),
and $ \Omega=3.0$ (c) for $N=4$. The dashed lines denote the
energy for the  total angular momentum $j=j_{max}=N/2$, which is
clearly higher the true ground-state energy. } \label{variation}
\end{center}
\end{figure}

The jump of $j$ in the $\lambda-\Omega$ plane exhibited in Fig.
\ref{firstorder} may be related to the first-order QPT in a finite
system. It is known that there should be a second-order QPT in this
systems without interatomic coupling. Since both the  first- and
second-order QPTs can be simply characterized by singularities of
the ground-state energy, we calculate ground-state energy and its
second derivative. These results together with the values of   $j$
are collected in Fig. \ref{variation} for three typical value of
$\Omega$ above the first-order QPT point at $\lambda=0$. For smaller
vale of $\Omega=2.2$, two jumps of $j$  indicates two first-order
QPTs, as shown in   Fig. \ref{variation}(a). The non-analytic
feature of the ground-state energy and sudden drop of its second
derivative at the two jump points give the evidence of the
first-order QPT. Above the last jump point, the ground-state energy
are continuous and its second derivative shows a smooth drop around
$\lambda_c$, demonstrating a sign of the second-order QPT.
$\lambda_c(\Omega, N)$ is also presented in Fig. \ref{firstorder},
which divides the $j=j_{max}$ regime into two parts. However, as
shown in   Fig. \ref{variation}(c), for larger value of $\Omega=3$,
there is only one jump of $j$ from $0$ to $j_{max}=N/2$ in the whole
coupling regime. Both non-analytic feature of the ground-state
energy and sudden drop of its second derivative suggest a
first-order QPT at this jump point. No sign of second-order QPT is
observed in this case. For the intermediate vale $\Omega=2.5$, two
first-order QPTs occur sequentially with the jump of $j$.  We think
that both the first and the second-order QPT occur in the same point
for $\Omega=2.5$ and $3.0$, and finally only the characteristic of
the stronger first-order QPT shows up at the last jump.

If we naively set the total angular momentum to be  $j_{max}=N/2$ as
in the pure Dicke model\cite{chenqh}, we can also calculate
"ground-state", which are also given in Fig. \ref{variation} with
dashed lines. It is really higher than the true the ground-state
energy. The continuous behavior in the whole coupling regime is
observed in these dashed curves, demonstrating the above interesting
feature in a finite system stems from the variation of the total
angular momentum.

Next, we will study the effect of the interatomic interaction on the
second-order QPT in the thermodynamical limit.  We focuss on the
question whether the interatomic interaction alters the universality
class  of the second-order QPT.

Recently,  the fidelity, a concept in quantum information theory,
has been extensively used to identify the QPTs in various many-body
systems from the perspective of the ground-state   wave
functions\cite {Quan,Zanardi,Cozzini,You,Gu,chens,zhou,Kowk}. In a
mathematical sense, the fidelity is the overlap between two ground
states where the transition parameters deviate slightly. However,
the fidelity depends on a arbitrary small amount of the transition
parameters, which in turn yields an artificial factor. Zanardi et al
\cite{Cozzini} introduced the Riemannian metric tensor and You et al
\cite{You} proposed the fidelity susceptibility (FS) to avoid this
problem independently.

\begin{figure}[tbp]
\centering
\includegraphics[width=8cm]{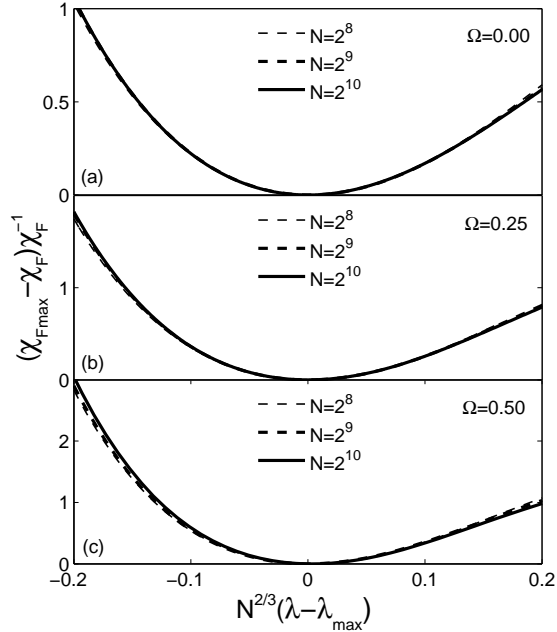}
\caption{Finite-size scaling of the average FS according to
Eq.(\ref{fs_scaling}) at the critical point for (a) $\Omega=0$, (b)
$\Omega=0.25$, and (c) $\Omega=0.5$. } \label{scaling}
\end{figure}

To analyze the QPT, we first  illustrate the scaling behavior of the
average FS. The finite-size scaling ansatz for the average FS  take
the form\cite{Kowk}
\begin{equation}
\frac{\chi _F^{\max }-\chi _F}{\chi _F}=f[N^\nu (\lambda -\lambda
_{\max })]\label{fs_scaling}
\end{equation}
where $\chi _F^{\max }$ is the value of average FS at the maximum
point $  \lambda _{\max }$, $f$ is the scaling function and $\nu $
is the correlation length critical exponent. This function should be
universal for large N in the second-order QPTs, which is independent
of the order parameter. As shown in Fig. \ref{scaling} an excellent
collapse in the critical regime is achieved   with the use of  $\nu
=2/3$  according to Eq.(\ref{fs_scaling}) in the curve for different
large size for three  values of $\Omega$. It is demonstrated that
$\nu $ is a universal constant and does not depended on the
parameter $\Omega$, suggesting the interatomic interaction does not
change the universality class.

\begin{figure}[tbp]
\centering
\includegraphics[width=8cm]{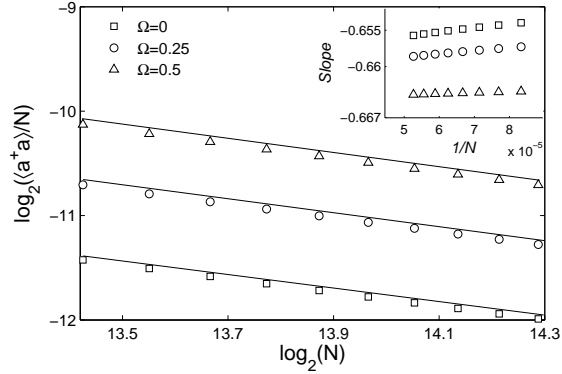}
\caption{Scaling of the order parameter $ \langle a^{\dagger}a
\rangle/N$ as a function of $N$ at the critical point for
$\Omega=0$,   $0.25$, and  $0.5$.  The inset shows the slope versus
$1/N$.} \label{order}
\end{figure}

We then perform the finite size scaling analysis on the order
parameter of the QPT, i.e. the  expectation value of  the photon
number per atom in the ground-state $ \langle a^{\dagger}a
\rangle/N$. In the thermodynamic limit, this quantity changes from
zero to finite value smoothly  when crossing the critical point. In
Fig. \ref{order}, we present this quantity as a function of $N$ for
different values of $\Omega$ in log-log scale. Derivatives of these
curves are plotted  in the inset. The exponent of the order
parameter is estimated to be ${-0.66\pm 0.01}$ for three values of
$\Omega$, provided another piece of the evidence that  the
interatomic interaction does not change the universality class.

\begin{figure}[tbp]
\centering
\includegraphics[width=8cm]{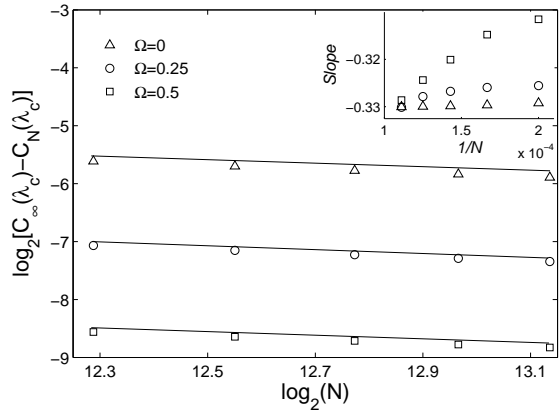}
\caption{Scaling of concurrence as a function of $N$ at the critical
point for $\Omega=0$,   $0.25$, and  $0.5$.  The  inset show the
slope versus $1/N$.} \label{entanglement}
\end{figure}

Finally, we calculate the scaled concurrence (entanglement) and
perform the corresponding finite size scaling analysis as in Ref.
\cite{chenqh,vidal,reslen}. The concurrence  quantifies the
entanglement between two atoms in the atomic ensemble after tracing
out over bosons.  In the thermodynamic limit, the scaled concurrence
at critical point $C_\infty (\lambda _c)$ can be easily determined
with the solution of  Eq. (11). For comparison, we can calculate the
quantity $ C_\infty (\lambda _c)-C_N(\lambda _c)$. In Fig.
\ref{entanglement}, we present this quantity as a function of $N$
for different values of $\Omega$ in log-log scale. Derivative of
these curves is presented in the inset, and the exponent of
concurrence is estimated to be ${-0.33\pm 0.005}$ for three values
of  $\Omega$. This again demonstrates that the universality class is
not altered by the interatomic interaction.

\section{Conclusion}

In summary, a finite number of  dipole-coupled two-level atoms   in
a single-mode resonant cavity is solved exactly with the use of
extended bosonic coherent in the numerically sense. A number ( the
maximum value is $N/2$)  of  first-order QPTs occur in this system
with $N$ coupled atoms, different from in the pure Dicke model. The
total angular momentum in the ground-state is altered with the
coupling constant and the interatomic interactions. The quantum
criticality is also studied in terms of the ground-state   fidelity,
order parameter and concurrence in very large systems up to $2^{10}$
or more.  The finite-size scaling analysis  for these observables
are performed. The corresponding scaling exponents obtained remains
unchanged with the interatomic integrations, demonstrating that the
university class is not changed.  It should be pointed out that all
eigenfunctions and eigenvalues obtained in  the modified Dicke model
on the large scale might also be used to  explore the mechanism for
the coherent radiation in one-dimensional Josephson arrays coupled
to a single-mode  cavity at both zero and finite temperatures, which
may be our future work.

\section*{ACKNOWLEDGEMENTS}

This work was supported by National Natural Science Foundation of
China, PCSIRT (Grant No. IRT0754) in University in China, National
Basic Research Program of China (Grant Nos. 2011CB605903 and
2009CB929104), Zhejiang Provincial Natural Science Foundation under
Grant No. Z7080203, and Program for Innovative Research Team in
Zhejiang Normal University.

$*$ Corresponding author. Email:qhchen@zju.edu.cn

\end{document}